\RequirePackage{fix-cm}
\documentclass[twocolumn]{svjour3}          
\smartqed  
\usepackage{graphicx}
\usepackage{amssymb}
\usepackage{parskip}
 \journalname{Soc. Netw. Anal. Min.}
\begin{document}

\title{Forensic Analysis of Phone Call Networks
}


\author{Salvatore Catanese  \and
        Emilio Ferrara			\and
        Giacomo Fiumara
}


\institute{S. Catanese, G. Fiumara	\at University of Messina, Dept. of Physics, Informatics Section, Via Ferdinando Stagno D'Alcontres, Salita Sperone, n. 31, Italy, 	
					\email{salvocatanese@gmail.com; giacomo.fiumara@unime.it}
					\and
          E. Ferrara \at University of Messina, Dept. of Mathematics, Via Ferdinando Stagno D'Alcontres, Salita Sperone, n. 31, Italy,
         	\email{emilio.ferrara@unime.it}					
}

\date{Received: date / Accepted: date}

\maketitle

\begin{abstract}
In the context of preventing and fighting crime, the analysis of mobile phone traffic, among actors of a criminal network, is helpful in order to reconstruct illegal activities on the base of the relationships connecting those specific individuals.
Thus, forensic analysts and investigators require new advanced tools and techniques which allow them to manage these data in a meaningful and efficient way.
In this paper we present \emph{LogAnalysis}, a tool we developed to provide visual data representation and filtering, statistical analysis features and the possibility of a temporal analysis of mobile phone activities.
Its adoption may help in unveiling the structure of a criminal network and the roles and dynamics of communications among its components.
By using \emph{LogAnalysis}, forensic investigators could deeply understand hierarchies within criminal organizations, for example discovering central members that provide connections among different sub-groups, etc.
Moreover, by analyzing the temporal evolution of the contacts among individuals, or by focusing on specific time windows they could acquire additional insights on the data they are analyzing. 
Finally, we put into evidence how the adoption of \emph{LogAnalysis} may be crucial to solve real cases, providing as example a number of case studies inspired by real forensic investigations led by one of the authors.
\end{abstract}

\section{Introduction}

The increasing usage of mobile phones in the everyday-life reflects also in their illicit adoption.
For example, mobile communication devices are exploited by criminal organizations in order to coordinate illegal activities, to communicate decisions, etc.
In order to prevent and fight crime, mobile communication service providers (according to the regulatory legislation of the State in which they operate) have to store for a given period all the data related to the phone traffic, in the shape of log files.
These logs contain information about phone calls, attempted calls, Short Message Service (SMS), Multimedia Messaging Service (MMS), General Packet Radio Service (GPRS) and Internet sessions.
Additional information could be inferred from traffic produced by Cell Global Identities (CGI)\footnote{CGI is a standard identifier for mobile phones cells which provides geographical positioning of mobile phones.} inside their areas.

The analysis of reports suppliedby mobile phone service providers makes it possible to reconstruct the network of relationships among individuals, such as in the context of criminal organizations.
It is possible, in other terms, to unveil the existence of criminal networks, sometimes called rings, identifying actors within the network together with their roles.
These techniques of forensic investigations are well-known, and are rooted in the Social Network Analysis (SNA). 
The structure of criminal networks could be efficiently formalized by means of graphs, whose nodes  represent actors of the criminal organizations (or, in our case, their mobile phones) and edges represent the connections among them (i.e., their phone communications).
The graph representation of data extracted from log files is a simple task, while its interpretation may result hard, when large volumes of data are involved.
In fact, it could become difficult to find anomalous values and models while browsing a large quantity of data.
Moreover, visual representations of a high number of individuals and connections easily become unreadable because of nodes and edges each other overlapping.
A powerful support comes from SNA, which provides methods to evaluate the importance of particular individuals within a network and relationships among them.
For example, SNA provides statistical algorithms that find those individuals/nodes in key positions and those acting as \emph{cohesive elements}.

In this work we present a novel tool we developed, named \emph{LogAnalysis}, for forensic visual statistical analysis of mobile phone traffic logs.
\emph{LogAnalysis} graphically represents the relationships among mobile phone users with a node-link layout.
It helps to explore the structure of a graph, measuring connectivity among users and giving support to visual search and automatic identification of organizations and groups within the network. 
To this purpose, \emph{LogAnalysis} integrates the graphical representation of networks with metrics and measures typical of SNA, in order to help detectives or forensic analysts to understand the structure of criminal associations while highlighting key members inside the criminal ring, and/or those members working as link among different associations, and so on.
Several statistical measures have been implemented and made available to the investigators, with a seamless integration with the visual part.
An additional feature is the possibility of analyzing the temporal evolution of the connections among actors of the network, for example focusing on particular time windows in order to obtain additional insights about the dynamics of communications before/during/after particular criminal events.
The main features of \textit{LogAnalysis} are described together with a number of case studies, inspired to a real criminal investigation brought by one of the authors, successfully solved also by exploiting features provided by \emph{LogAnalysis}.

\section{Related work}

Law enforcement and intelligence agencies frequently face the problem of extracting information from large amounts of raw data coming from heterogeneous sources, among which are phone calls printouts.
In recent years, a growing number of commercial software has been developed that employ analytical techniques of visualization to help investigations. 
In the following we briefly describe, at the best of our knowledge, the most successful among them.

Analysts Notebook from i2 Inc.\footnote{i2 - Analysts Notebook. http://www.i2inc.com/} provides a semantic graph visualization to assist analysts with investigations. Nodes in the graph are entities of semantic data types such as persons, events, organizations, bank accounts, etc. 
While the system can import text files and do automatic layout, its primary application appears to be helping analysts in manually creating and refining case charts.

The COPLINK system \cite{Chen2003} and the related suite of tools has a twofold goal: to ease the extraction of information from police case reports and to analyze criminal networks.
A conceptual space of entities and objects is built exploiting data mining techniques in order to help in finding relations between entities.
It also provides a visualization support consisting of a hyperbolic tree view and a spring-embedder graph layout of relevant entities.
Furthermore, COPLINK is able to optimize the management of information exploited by police forces integrating in a unique environment data regarding different cases.
This is done in order to enhance the possibility of linking data from different criminal investigations to get additional insights and to compare them in an analytic fashion.

\texttt{TRIST} \cite{jonker2005information} allows analysts to formulate, refine, organize and execute queries over large document collections. 
Its user interface provides different perspectives on search results including clustering, trend analysis, comparisons, and difference. 
Information retrieved by \texttt{TRIST} then can be loaded into the \texttt{SANDBOX} system \cite{wright2006sandbox}, an analytical sense-making environment that helps to sort, organize, and analyze large amounts of data. 
The system offers interactive visualization techniques including gestures for placing, moving, and grouping information, as well as templates for building visual models of information and visual assessment of evidence.
Similarly to COPLINK, TRIST is optimized to query large databases and to analytically compare results.

Differently from COPLINK and TRIST, \emph{LogAnalysis} adopt a different approach, which is not based on querying data but it relies on full visual presentation and analysis of such information represented by means of network graphs.
The strength of our tool is the adoption of several interactive layout techniques that highlight different aspects and features of the considered networks and it allows the inspection of elements (nodes and edges) that constitute the network itself.

Another remarkable tool is \texttt{GeoTime} \cite{Kapler2004}, that visualizes the spatial interconnectedness of information over time overlaid onto a geographical substrate. 
It uses an interactive 3D view to visualize and track events, objects, and activities both temporally and geo-spatially. 
One difference between \texttt{GeoTime} and \emph{LogAnalysis} is that the feature regarding the spacial dependency of data is not yet allowed by our tool, and this makes \texttt{GeoTime} a useful addition to \emph{LogAnalysis} for such type of investigations.
On the other hand, the functionalities provided by \emph{LogAnalysis} in terms of analysis of temporal dependencies of data improve those provided by \texttt{GeoTime}, as highlithed in Section \ref{sub:time-filtering}--\ref{sub:stacked-histograms}.

As an example of the various general-purpose tools for analyzing social networks (differently from tools specifically designed to investigate telecom networks), we mention NodeXL \cite{smith2009analyzing}, an extensible toolkit for network overview, discovery and exploration implemented as an add-on to the Microsoft Excel 2007/2010 spreadsheet.
NodeXL is open source and was designed to facilitate learning the concepts and methods of Social Network Analysis with visualization as a key component.
It integrates metrics, statistical methods, and visualization to gain the benefit of all the three approaches.
As for the usage of network metrics to assess the importance of actors in the network, NodeXL shares a paradigm similar to that we adopted in \emph{LogAnalysis}, although it lacks of all the relevant features of our tools related to the temporal analysis of the networks. 

Regarding those researches that apply Social Network Analysis to relevant topics related to this work, recently T. von Landesberger et al. \cite{Landesberger2011} surveyed the available techniques for the visual analysis of large graphs. 
Graph visualization techniques are shown and various graph algorithmic aspects are discussed, which are useful for the different stages of the visual graph analysis process. 
In this work we received a number of challenges proposed by \cite{Landesberger2011}, trying to address for example the problem of large-scale network visualization for ad-hoc problems (in our case, to study phone telecom networks).

Also the analysis of phone call networks has been a subject of intensive study.
Mellars \cite{mellars2004forensic} investigated the principal ways a phone call network operates and how data are processed.
Particular attention has been given to the methodology of investigation of data about the phone activity that it is possible to collect directly from the devices.

More recently, different works \cite{palla2007quantifying,onnela2007structure,Onnela2007b,blondel2008fast} used mobile phone call data to examine and characterize the social interactions among cell phone users. 
They analyze phone traffic networks consisting of the mobile phone call records of million individuals.

In details, in \cite{onnela2007structure,Onnela2007b} the authors present the statistical features of a large-scale Belgian phone call network constituted by 4.6 millions users and 7 millions links. 
That study highlights some features typical of large social networks \cite{ferrara2011topological} that characterize also telecom networks, such as the fission in small clusters and the presence of strong and weak ties among individuals. 
In addition, in \cite{palla2007quantifying} the authors discuss an exceptional feature of that network, which is the division in two large communities corresponding to two different language users (i.e., English and French speakers of the Belgian network).

The community structure of phone telecom networks has been further investigated in \cite{blondel2008fast}.
The authors exploited an efficient community detection algorithm called \emph{Louvain method} \cite{blondel2008fast,ferrara2011generalized} to assess the presence of the community structure and to study its features, in a large phone network of 2.6 millions individuals.

In conclusion, during the latest years Eagle et al. \cite{eagle2008mobile,eagle2009inferring} investigated the possibility of inferring a friendship social network based on the data from mobile phone traffic of the same individuals.
This problem attracted the attention of other recent studies \cite{candia2008uncovering,sundsoy2010product}, particularly devoted to understand the dynamics of social connections among individuals by means of mobile phone networks.

\subsection{Contribution of this work}
\emph{LogAnalysis} has been originally presented in a preliminary version during late 2010 \cite{catanese2010visual} and has received a positive critique by the research community of \emph{forensic analysts} and \emph{social network analysts}.

We argue that the further developments of this tool have increased its potential and performance.
In particular, the research direction that we are following with \emph{LogAnalysis} is devoted to include the possibility of analyzing temporal information from \emph{phone call networks}, and the tool has been specifically optimized to study \emph{mobile phone telecom networks}, whose analysis has attracted relevant research efforts in the recent period \cite{saravanan2011analyzing}.
Additional efforts have been carried out so that to improve the possibilities provided by \emph{LogAnalysis} to unveil and study the community structure of the networks, whose importance has been assessed during latest years in a number of works \cite{porter2009communities,gilbert2010communities}, by means of different community detection techniques \cite{fortunato2010community,ferrara2011generalized,coscia2011classification}.

Our tool introduces a number of novelties with respect to similar platforms existing as to date.
In detail, \emph{LogAnalysis} primarily differs from the systems described above as we focused on the visual representation of the relationships among entities in phone calls. 
We adopted different state-of-the-art view layouts for promoting fast exploration and discovery of the analyzed networks. 

Furthermore, our tool provides a system model which aims at improving the quality of the analysis of social relationships of the network through the integration of visualization and SNA-based statistical techniques, which is a relevant topic in the ongoing research in Social Network Analysis \cite{scott2011social}.

To this purpose, \emph{LogAnalysis} has been assessed as an invaluable support during real investigations carried out by professional \emph{forensic analysts}, in particular in the context of analyzing large-scale \emph{mobile telecom networks} exploited for criminal purposes.

One of the merits of this work, in fact, is to analyze several different real-world use cases inspired by forensic investigations carried out by one of the authors.
During these investigations, \emph{LogAnalysis} has been exploited to examine the structural features of criminal phone call networks with a systematic methodology adopting a unique tool, differently from previous cases in which a combination of different SNA-based and digital forensic tools had to be adopted to reach similar results.
Some relevant information about the usage of \emph{LogAnalysis} in the context of real-world investigations have been reported in this work.
As the best of our knowledge, this is the first work to present critical information from real forensic investigations in mobile phone call networks, dealing with real data acquired from actual criminal cases.
As a relevant fact, we provide with some clues that support our claim about the advantages of adopting \emph{LogAnalysis} to unveil possible criminal connections among actors of mobile telecom networks.

\section{Analysis of Mobile Phone Traffic Networks}
The relationships established by means of phone calls may be explored using different techniques and approaches.
Sometimes, forensic analysis relates to phone traffic made by International Mobile Subscriber Identity (IMSI)\footnote{IMSI is a unique number associated with all GSM and UMTS network mobile phone users. It is stored in the SIM inside the phone and is sent by the phone to the network.} and by International Mobile Equipment Identity (IMEI)\footnote{IMEI is a unique 17 or 15 digit code used to identify an individual mobile station to a GSM or UMTS network.}.
Detectives generally distinguish three main types of analysis of phone traffic logs: 
i) \textit{relational}, in order to show links (and hence acquaintance) among individual users; 
ii) \textit{spatial}, helpful to show geographical displacements of a mobile phone in order to assess location of an individual before, during and after a crime has been committed and, 
iii) \textit{temporal}, useful to discover, for example, at what time a phone call has been made or a SMS has been sent, which contacts were involved in a phone conversation or how long an Internet connection lasted.
\emph{LogAnalysis} provides some tools to investigate relational and temporal aspects of phone call networks.

\begin{figure}
	\centering
	\includegraphics[width=3.3in]{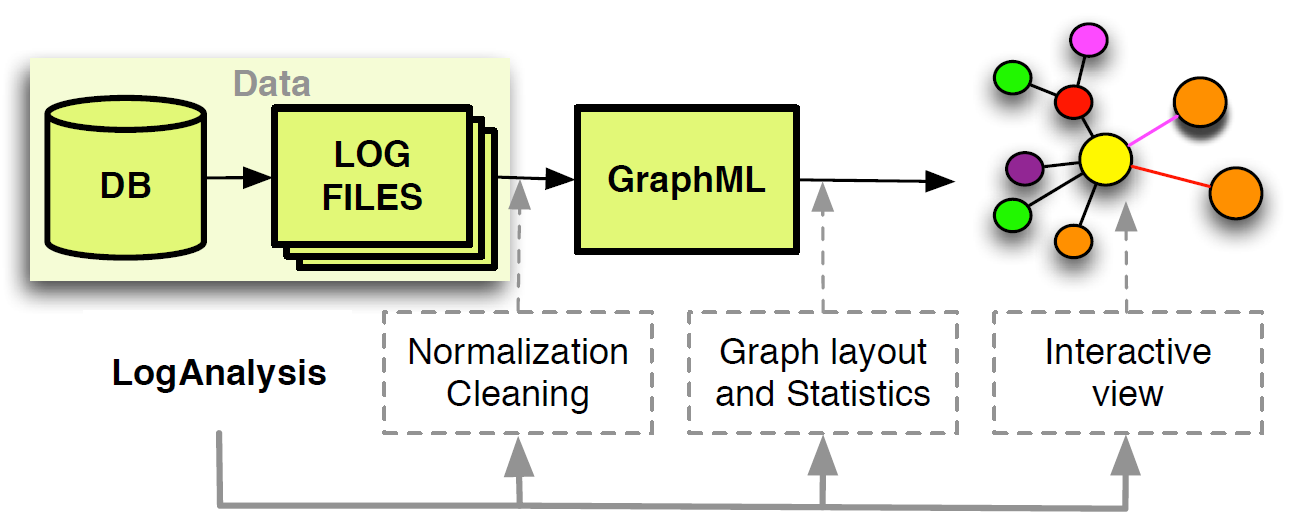}
	\caption{Architecture of \emph{LogAnalysis}.}
	\label{fig:architecture}
\end{figure}

The architecture of \emph{LogAnalysis}, shown in Figure \ref{fig:architecture}, is designed by extensible levels: i) import of data provided by informative systems of mobile phone service providers (usually, under the form of \emph{textual log files}); ii) conversion of data to the GraphML\footnote{http://graphml.graphdrawing.org/} format, a structured XML format, more suitable for graphical representation and portability among several different graph drawing applications; iii) visualization and dynamic exploration of the obtained mobile phone traffic network.

An example of the usage of this tool is the research of particular elements in the network.
To this purpose, it is possible to visually discover subsets and gangs (or rings), by measuring their cohesion in terms of the density of internal connections.
Thus, from the overall structure of the network are extracted those elements of interest for investigations.
In fact, some nodes are prominent due to their high degree of connection with others, other nodes for their strategic position of centrality in terms of connections, etc.
A number of complete case studies has been analyzed in Section \ref{results}, describing some features provided by \emph{LogAnalysis} that have been exploited in the context of a real-world investigation.

\subsection{Implementation}
Our system is implemented in Java and integrates several open-source toolkits. 
In particular, \emph{Prefuse}\footnote{http://prefuse.org/} provides the underlying node-link data structures and has been used to support the dynamic exploration of networks, according to force-directed and radial models, and to identify communities.
\emph{JUNG}\footnote{http://jung.sf.net/} has been used to implement some of the SNA ranking algorithms, for the computation and visualization of the shortest path(s) connecting a pair of nodes, and for the network visualizations and clustering.

\subsection{Data Import}
In the context of real-world investigations, mobile phone service providers, upon request by judiciary authorities, release data logs, normally in textual file format, with space or tab separation (CSV format).
A typical log file contains, at least, the values shown in Table \ref{tab:fileformat}.

Similarly, information about owners of SIM cards, dealers of SIM cards and operations like activation, deactivation, number portability are provided by the service providers as additional material in order to ease and support the investigation activities.
Log file formats produced by different companies are heterogeneous.
\emph{LogAnalysis}, first of all, parses these files and converts data into GraphML format.
It is an XML valid and well-formed format, containing all nodes and weighted edges, each weight representing the frequency of phone calls  between two adjacent nodes.
GraphML has been adopted both because of its extensibility and ease of import from different SNA toolkits and graph drawing utilities.

\subsection{Data Normalization/Cleaning}
Data clean-up usually means the deletion of redundant edges and nodes.
This step is very importante since datasets often contain redundant information, that crowds graph visualization and biases statistical measures.
In these circumstances, redundant edges between the same two nodes are collapsed and a coefficient -- i.e., a edge weight -- is attached, which expresses the number of calls.
Our tool normalizes data after reading and parsing log files whichever format they have been provided among the standard formats (i.e., \emph{fixed width text}, \emph{delimited}, CSV, and more) used by mobile service providers.

\begin{table}
	\centering 	\small
	\begin{tabular}{|l|l|} 
		\hline
		\textbf{Field}&\textbf{Description}\\ 
		\hline
		\texttt{IMEI} & IMEI code MS\\ 
		\hline
		\texttt{called} & called user\\ 
		\hline
		\texttt{calling} & calling user\\ 
		\hline
		\texttt{date/time start} & date/time start calling (GMT)\\ 
		\hline
		\texttt{date/time end} & date/time end calling (GMT)\\ 
		\hline
		\texttt{type} & sms, mms, voice, data etc. \\ 
		\hline
		\texttt{IMSI} & calling or called SIM card\\ 
		\hline
		\texttt{CGI} & Lat. long. BTS company\\
		\hline
	\end{tabular}
	\caption{An example of the structure of a log file.}
	\label{tab:fileformat}
\end{table}

\section{Eyes on some Features of \emph{LogAnalysis}}
In this section we put into evidence some of the main features of \emph{LogAnalysis} that have been inspired both by forensic analysis and the social network analysis.
In particular, in Section \ref{sub:data} we point out those data exploration features provided by our tool.
Subsequently, in Section \ref{sub:centrality} we discuss the role and functioning of a set of centrality measures implemented in \emph{LogAnalysis} that can be exploited to assess the importance of actors of mobile telecom networks. 
Furthermore, in Section \ref{sub:layout} the layout models adopted in our tool are described, focusing on the novelties introduced by \emph{LogAnalysis} in respect to general purpose SNA tools.

\subsection{Data Exploration} \label{sub:data}
The main goal of \emph{LogAnalysis} is to support the forensic detectives into the exploration of data provided by mobile phone service providers about the phone traffic activity of particular individuals of interest for the forensic investigations.
This support is given by means of an interactive visual representation of the phone traffic network.
To this purpose, individuals are identified by means of their phones and are represented by nodes of a graph. 
The phone calls, instead, represent the interactions among actors and for this reason they are captured as the edges of the same graph.
More formally, the structure of phone traffic is described in terms of directed graphs $G=(V,E)$ where $V$ is the set of telephone numbers (nodes) and $E$ is the set of calls (edges) among the nodes. 
The edges, directed and weighted, show the direction (incoming or outgoing) and the number of phone calls between the various pairs of adjacent nodes.

\emph{LogAnalysis} is able to manage phone traffic networks up to hundred thousands elements and log files up to millions entries.
However, in our experience, a meaningful interactive visual representation of these data is viable analyzing networks up to some thousands of elements.
To this purpose, in Section \ref{results} we describe a number of case studies inspired by real investigations whose network includes thousands elements, and in which \emph{LogAnalysis} played a fundamental role in the successful conclusion of the investigation.

In detail, one of the most useful features of \emph{LogAnalysis} is that it is able to identify and visually put into evidence those actors in the network that play a crucial role in the communication dynamics. 
This is done by exploiting the centrality measures provided by the Social Network Analysis (described in the next section).
On the other hand, a visual layout only could not be sufficient to put into evidence all the required information.
For example, different visual representation would help detectives to reach additional insights about data, the dynamics of the phone traffic network and the activities of the actors of the network. 
For this reason, \emph{LogAnalysis} provides different interactive visual representations, by adopting several algorithms.

\subsection{Centrality Measures} \label{sub:centrality}
\emph{LogAnalysis} takes into account the concept of \emph{centrality measure} to highlight actors that cover relevant roles inside the analyzed network.
Several notions of centrality have been proposed during the latest years in the context of Social Network Analysis. 

There are two fundamentally different class of centrality measures in communication networks. 
The first class of measures evaluates the centrality of each node/edge in a network and is called point centrality measure. 
The second type is called graph centrality measure because it assigns a centrality value to the whole network.
These techniques are particularly suited to study phone traffic and criminal networks.

In detail, in \emph{LogAnalysis} we adopted four point centrality measures (i.e., \emph{degree}, \emph{betweenness}, \emph{closeness} and \emph{eigenvector} centrality), to inspect the importance of each node of the network. 

The set of measures provided in our tool is a selection of those provided by Social Network Analysis \cite{wasserman1994social}.
It could be not sufficient to solve any possible task in phone call network analysis. 
In fact, for particular assignments it could yet be necessary to use additional tools in support to \emph{LogAnalysis} and in further evolutions we plan to incorporate new centrality measures \cite{ferrara2011novel,abdallah2011generalizing} if necessary.   

For each centrality measure, the tool gives the possibility, to rank the nodes/edges of the network according to the chosen criterion.
Moreover, \emph{LogAnalysis} allows to select those nodes that are central, according to the specified ranking, highlighting them and putting into evidence their relationships, by exploiting the node-link layout techniques (discussed in the following). 
This approach makes it possible to focus the attention of the analysts on specific nodes of interest, putting into evidence their position and their role inside the network, with respect to the others.

In the following we formally describe the centrality measures used in \emph{LogAnalysis}. 

They represent the centrality as an indicator of the activity of the nodes (degree centrality), of the control on other nodes (betweenness centrality), of the proximity to other nodes (closeness centrality) and of the influence of a node (eigenvector centrality).

\subsubsection{Degree centrality}
The degree centrality of a node is defined as the number of edges adjacent to this node.
For a directed graph $G=(V,E)$ with $n$ nodes, we can define the
in-degree and out-degree centrality  measures as

\begin{equation}
	C_D(v)_{in}=\frac{d_{in}(v)}{n-1}, \quad C_D(v)_{out}=\frac{d_{out}(v)}{n-1}
	\label{eq:degree}
\end{equation}

where $d_{in}(v)$ is the number of incoming edges adjacent to the node $v$, and $d_{out}(v)$ is the number of the outgoing ones.

Since a node can at most be adjacent to $n - 1$ other nodes, $n - 1$ is the normalization factor introduced to make the definition independent on the size of the network and to have $0 \le C_D(v) \le 1$.

In and out-degree centrality indicates how much activity is going on and the most active members.
A node with a high degree can be seen as a hub, an active nodes and an important communication channel.

We chose to include the degree centrality for a number of reasons.
First of all, is calculation is computationally even on large networks.
Furthermore, in the context of phone call networks it could be interpreted as the chance of a node for catching any information traveling through the network.

Most importantly, in this type of directed networks, high values of in-degree are considered a reliable indicator of a form of popularity/importance of the given node in the network; on the contrary, high values of out-degree are interpreted as a form of gregariousness of the given actor in respect to the contacted individuals.

\subsubsection{Betweenness centrality}
The communication between two non-adjacent nodes might depend on the others, especially on those on the paths connecting the two nodes. 
These intermediate elements may wield strategic control and influence on many others.

The core issue of this centrality measure is that an actor is central if she lies along the shortest paths connecting other pairs of nodes.
The betweenness centrality of a node $v$ can be defined as

\begin{equation}
	B_C(v)=\sum_{s\ne{v}\ne{t}}\frac{\sigma_{st}(v)}{\sigma_{st}}
	\label{eq:betweenness}
\end{equation}

where $\sigma_{st}$ is the number of shortest paths from $s$ to $t$ and $\sigma_{st}(v)$ is the number of shortest paths from $s$ to $t$ that pass through a node $v$.

The importance of the betweenness centrality regards its capacity of identifying those nodes that vehiculate information among different groups of individuals.

In fact, since its definition due to Freeman \cite{freeman1977set} the betweenness centrality has been recognized as a good indicator to quantify the ability of an actor of the network to control the communication between other individuals and, specifically for this reason it has been included in \emph{LogAnalysis}.

In addition, it has been exploited by Newman \cite{newman2004fast} to devise an algorithm to identify communities within a network.
Its adoption in the phone traffic networks is crucial in order to identify those actors that allow the communication among different (possibly criminal) groups.

\subsubsection{Closeness centrality}
Another useful centrality measure that has been adopted in \emph{LogAnalysis} is called \emph{closeness centrality}.
The idea is that an actor is central if she can quickly interact with all the others, not only with her first neighbors \cite{Newman_2005}.
The notion of closeness is based on the concept of shortest paths (geodesic) $d(u,v)$, the minimum number of edges traversed to get from $u$ to $v$. The closeness centrality of the node $v$ is define as

\begin{equation}
	C_C(v)=\frac{1}{\sum_{u \in{V}} d(u,v)}
	\label{eq:closeness-centrality}
\end{equation}

Such a measure is meaningful for connected graphs only, assuming that $d(u,v)$ may be equal to a finite value.

In the context of criminal networks, this measure highlights entities with the minimum distance from the others, allowing them to pass on and receive communications more quickly than anyone else in the organization.
For this reason, the adoption of the closeness centrality is crucial in order to put into evidence inside the network, those individuals that are closer to others (in terms of phone communications).

In addition, high values of closeness centrality in such type of communication networks are usually regarded as an indicator of the ability of the given actor to quickly spread information to all other actors of the network.
For such a reason, the closeness centrality has been selected to be included in the set of centrality measures adopted by \emph{LogAnalysis}.

\subsubsection{Eigenvector centrality}
Another way to assign the centrality to an actor of the network in \emph{LogAnalysis} is based of the idea that if a node has many central neighbors, it should be central as well. 
This measure is called \emph{eigenvector centrality} and establishes that the importance of a node is determined by the importance of its neighbors.

The eigenvector centrality of a given node $v_i$ is 
\begin{equation}
	C_E(v_i) \propto \sum_{u \in N_i}{A_{ij}C_E(u)}
	\label{eq:eigenvector-centrality}
\end{equation}

where $N_i$ is the neighborhood of the given node $v_i$, and $x \propto Ax$ that implies $Ax = \lambda x$.
The centrality corresponds to the top eigenvector of adjacency matrix $A$.

In the context of telecom networks, eigenvector centrality is usually regarded as the measure of influence of a given node.
High values of eigenvector centrality are achieved by actors who are connected with high-scoring neighbors, which in turn, inherited such an influence from their high-scoring neighbors and so on.

This measure well reflects an intuitive important feature of communication networks that is the influence diffusion and for such a reason we decided to include the eigenvector centrality in \emph{LogAnalysis}.

\subsubsection{Clustering coefficient (transitivity)}

The clustering (or transitivity) coefficient of a graph measures the degree of interconnectedness of a network or, in other words, the tendency of two nodes that are not adjacent but share an acquaintance, to get themselves in contact.
High clustering coefficients mean the presence of a high number of triangles in the network.

The local clustering coefficient $C_i$ for a node $v_i$ is the number of links among the nodes within its neighborhood divided by the number of links that could possibly exist among them

\begin{equation}
	C_i=\frac{|\{e_{jk}\}|}{k_i(k_i-1)}:v_j,v_k \in N_i,e_{jk} \in E
	\label{eq:clustering}
\end{equation}

where the neighborhood $N$ of a node $v_i$ is defined as $N_i=\{v_j:e_{ij} \in E \wedge e_{ji} \in E\}$, while $k_i(k_i-1)$ is the number of links that could exist among the nodes within the neighborhood.

In is well-known in the literature \cite{wasserman1994social} that communication networks show high values of clustering coefficient since they reflect the underlying social structure of contacts among friends/acquaintances.
Moreover, high values of local clustering coefficient are considered a reliable indicator of of nodes whose neighbors are very well connected and among which a substantial amount of information may flow.
For such a reason, \emph{LogAnalysis} provides the possibility of computing both the global clustering coefficient for any given phone call network and the local clustering coefficient of any given node.

\subsection{Layout Algorithms} \label{sub:layout}
In this section we introduce the strategies of interactive visual representation of the phone traffic networks adopted in \emph{LogAnalysis}.
In detail, the graphical representation of phone relationships in \textit{LogAnalysis} exploits features provided by two well-known toolkits, \emph{Prefuse} and \emph{JUNG}.

\subsubsection{Force-directed Model}
\begin{figure*}[!ht]	
	\centering
	\includegraphics[width=2\columnwidth]{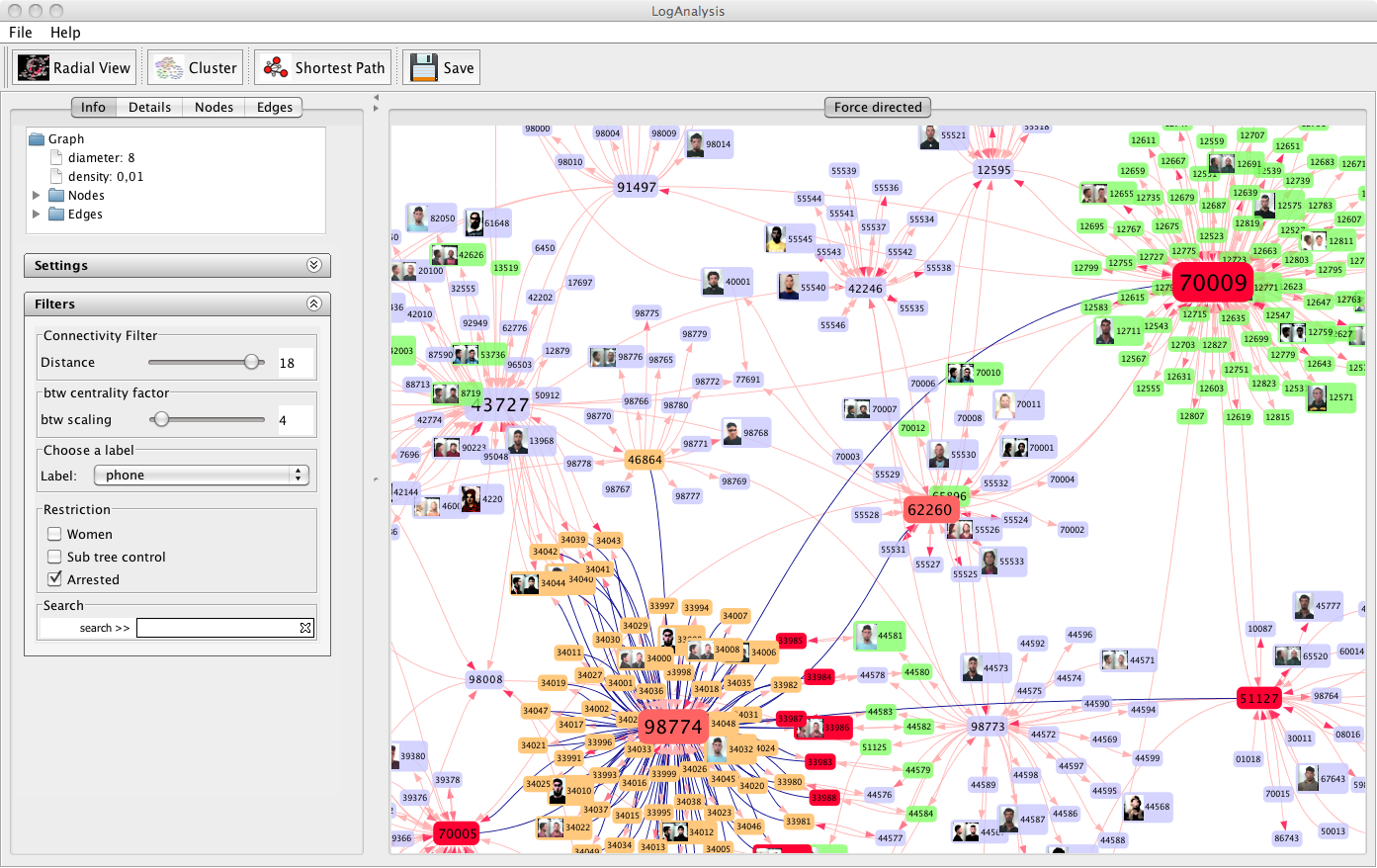}%
	\caption{Example of the force-directed visualization model. It is possible to put into evidence that groups of nodes repeal each other with respect to the weights of the edges connecting them.}%
	\label{fig:loganalysis}%
\end{figure*}

The main visual representation strategy adopted in \emph{LogAnalysis} is the called \emph{force-directed model}.
It is computed using the Fruchterman-Reingold algorithm \cite{fruchterman1991graph}, in which nodes repel each other and edges act as springs.
The consequent displacement of nodes and links shows users clustered in groups which can be identified on the base of their increase of connectivity.
The Barnes-Hut algorithm \cite{barnes1986hierarchical} simulates a N-body repulsive system in order to continuously update positions of elements.
Optimization of visualization is interactively obtained  by modifying parameters relative to the tensions of springs.
Nodes with minor connectivity have greater tension, resulting in a displacement of the elements of a group in \emph{orbital} position with respect to the central group.
In \emph{LogAnalysis} it is possible to modify different parameters, for example spring constant of force, gravitation force and viscosity/drag of forces.
In Figure \ref{fig:loganalysis} it is possible to appreciate an example of the force-directed visualization model.

\subsubsection{Edge Betweenness Clusterer}
We have found that the Fruchterman-Reingold layout in conjunction with the Edge Betweenness Clusterer \cite{girvan2002community} allows the interactive discovery of groups (henceforth, called \emph{clans}) existing inside the network and those individuals acting as links among groups (hereafter, called \emph{referents}).
This feature is crucial because it allows to forensic analysts to highlight with low efforts those clans whose activity may be suspect inside the phone call network. 
Moreover, it leads to additional insights in particular regarding the interconnection of these referents among each other and among clans.

More generally, the Edge Betweenness Clusterer, introduced by Newman \cite{newman2004fast}, is instrumental in the discovery of groups (called \emph{communities} in Social Network Analysis). 
This algorithm takes into account the weights of the edges in the network. 
In the particular scenario of the phone traffic networks, the concept of weights has already been defined as the number of phone communications among individuals.
To highlight the clans, \emph{LogAnalysis} exploits this technique according to a specific visualization strategy, called \emph{visual aggregation}.

\subsubsection{Visual Aggregation}
\emph{LogAnalysis}  adopts two algorithms to detect aggregations inside the network which represents the phone traffic.
The first algorithm called \emph{Edge Betweenness Clusterer} has been previously introduced.
To this purpose, instead of regarding the \emph{betweenness centrality} associated to a node, we consider the \emph{betweenness centrality} of an edge, which is defined as the number of shortest paths connecting pairs of nodes traversing it.

In the context of the visual aggregation, once the ranking of the edges is calculated, the algorithm simulates the deletion of those edges with the highest centrality, one by one, obtaining the effect of clustering the network in different groups (i.e., clans) that are weakly coupled each other but densely interconnected within them.
The functioning of this algorithm is based on the intuition that edges with high centrality connect groups characterized by high interconnectedness among their members and low outgoing connections.
The Edge Betweenness Clusterer has been proved to work well in the context of social networks.
To the best of our knowledge, this is the first attempt to adopt this strategy to identify clans inside phone traffic networks.

\begin{figure}
	\centering
	\includegraphics[width=\columnwidth]{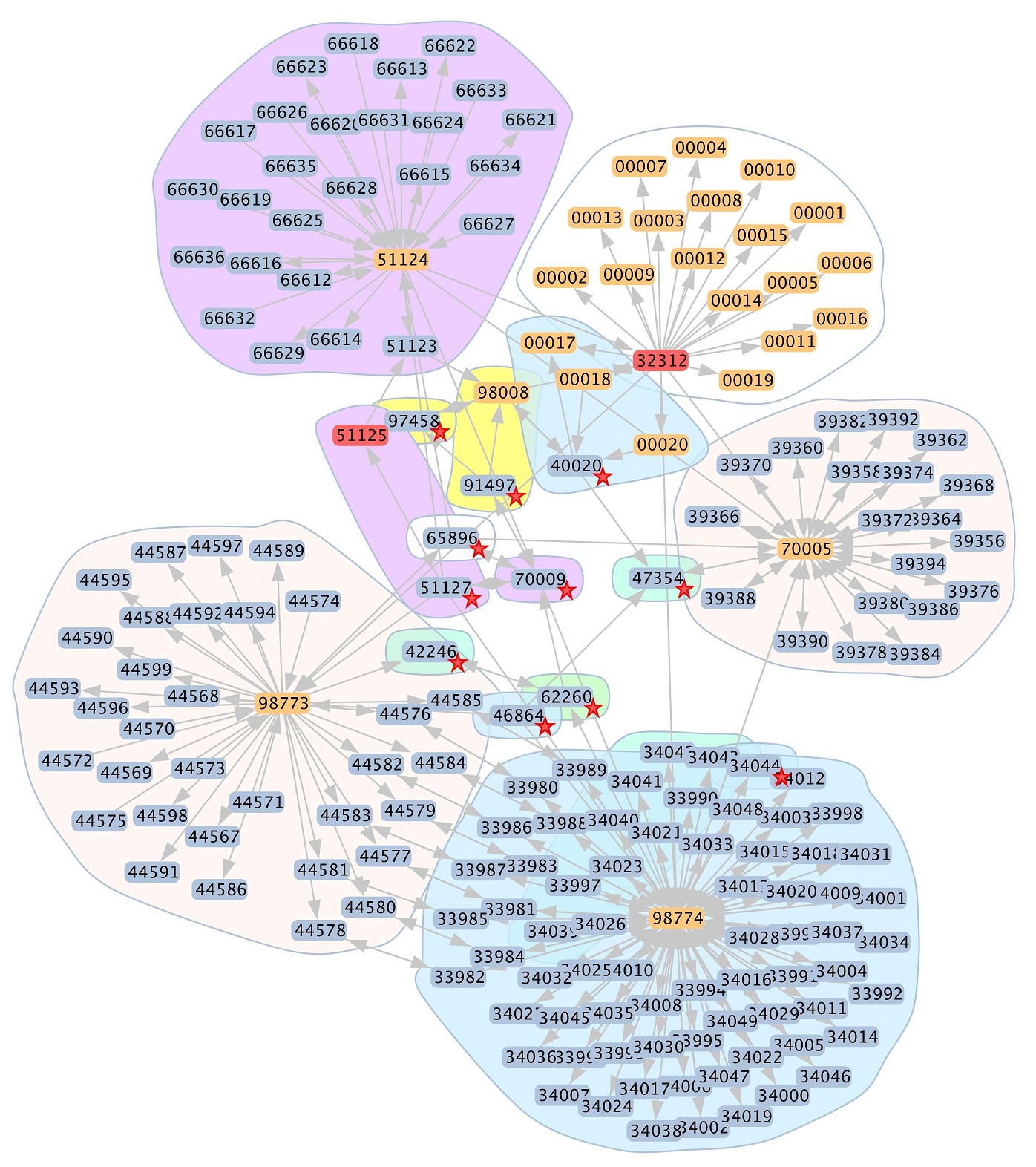}
	\caption{Example of visual aggregation layout. It is based on the Edge Betweenness Clusterer to divide the network in different clans on the base of the interactions among members.}
	\label{fig:visual-aggregation}
\end{figure}

The second algorithm, known as \emph{Newman's community identification algorithm} \cite{newman2004fast}, is a variant of the hierarchical agglomerate clustering (it is also adopted in Vizster \cite{heer2005vizster}). 
Regardless the adopted algorithm, \emph{LogAnalysis} visually presents the identified clans by surrounding all members with a translucent convex hull (see Figure \ref{fig:visual-aggregation}).

By expanding the action of filtering one can obtain interesting visualizations.
Groups comprised of a single node (i.e., monadic clusters) which satisfy the filtering condition are compressed and shown as a star.

Moreover, interactive cluster discovery is available in \emph{LogAnalysis}.
Users can suppress an arbitrary number of edges to discover strategic groups, together with links among groups.
Target edges are chosen according to the algorithm known as Edge Betweenness Clusterer.
Labels of nodes belonging to the same cluster bear the same color.
Different colors identify elements not belonging to a cluster.

\subsubsection{Radial Tree Layout}
The third layout algorithm introduced in \emph{LogAnalysis} is called Radial Tree. 
It allocates the elements of a graph in radial positions and defines several levels upon concentric circles with progressively increasing radii.
The algorithm developed by Ka Ping Yee et al. \cite{yee2001animated} also puts nodes in radial positions but gives the possibility of varying positions while preserving both orientation and order.

According to that technique, a selected element is placed at the center of the canvas and all the other nodes are subsequently placed upon concentric circles with radii increasing outwards.
This visualization strategy is instrumental in the context of the forensic analysis because it allows to focus the attention of detectives on a suspect, and to have a close look to its connections.

\begin{figure*}[!ht]%
	\centering
	\includegraphics[width=2\columnwidth]{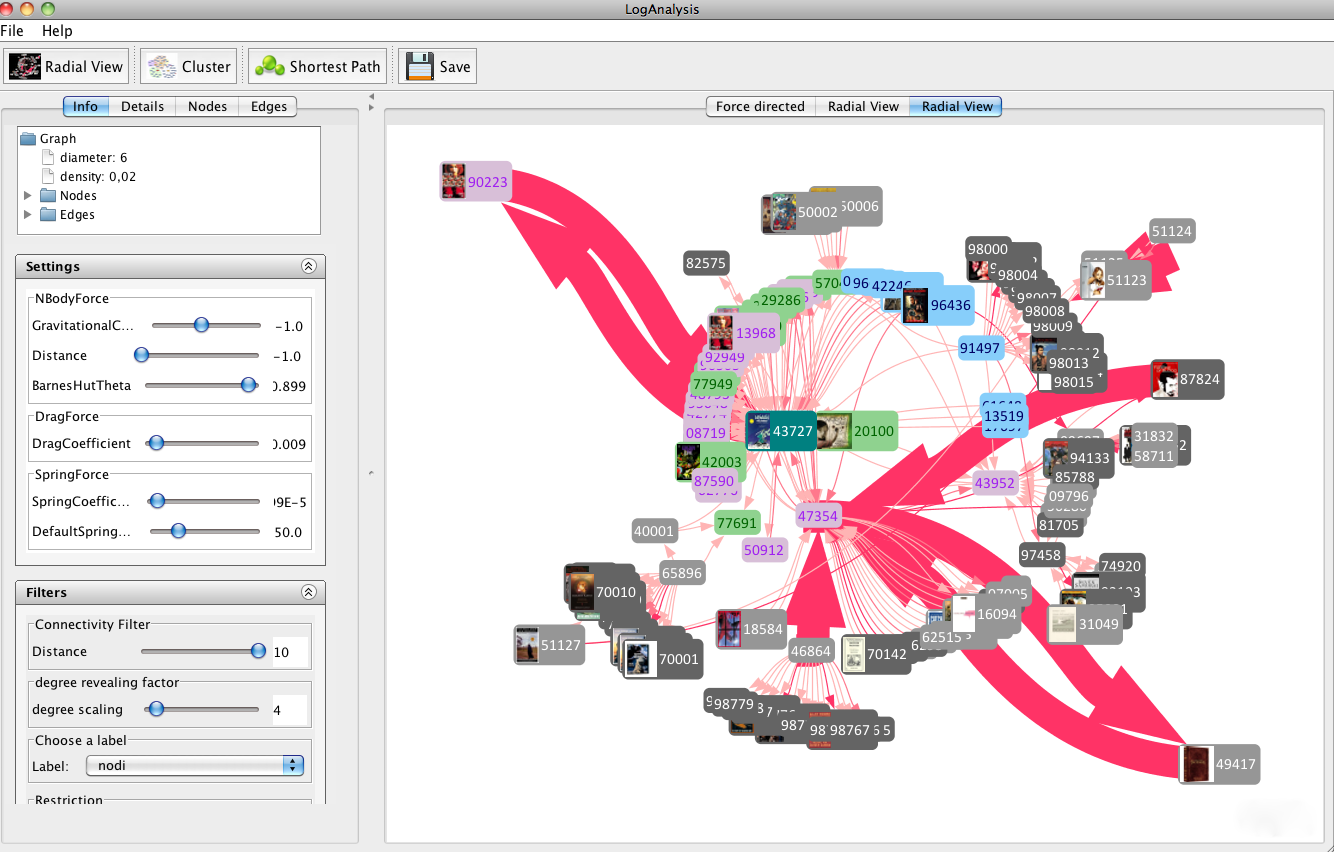}%
	\caption{Example of Radial View layout. A node specified by the analyst is put into the center of the visualization. In addition, nodes represented with different colors belong to different groups (i.e., clans).}%
	\label{fig:radial}%
\end{figure*}

The interactive visualization by using the Radial Tree layout is shown in Figure \ref{fig:radial}.
The interface supports filtering and searching elements within the network; a forensic analyst could select a specific node, which is placed at the center of the canvas.
Nodes lying on the circumference of concentric circles, centered on that node, could be also progressively displaced from the selected one.
Moreover, edges are visualized by using different thickness, calculated with respect to the number of calls among the given connected nodes.
The overlapping of nodes may be avoided by superimposing a force-directed visualization to the radial tree algorithm.

\begin{figure*}[!ht]%
	\centering
	\includegraphics[width=2\columnwidth]{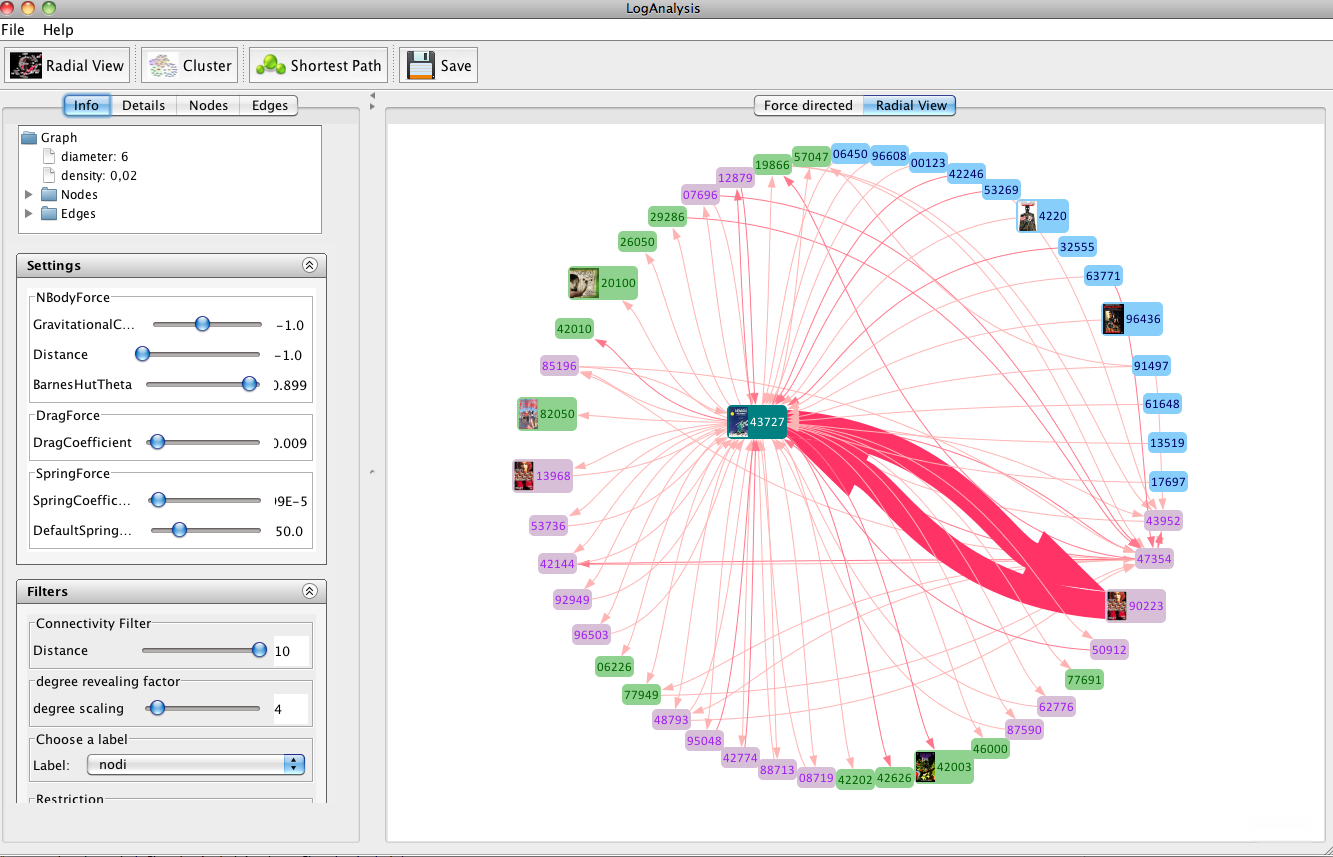}%
	\caption{Example of exploded Radial View layout. A selected node is put into the center and the neighborhood is presented in a radial layout. Nodes with different colors belong to different groups.}%
	\label{fig:radial-exploded}%
\end{figure*}

A useful extension that we implemented, shown in Figure \ref{fig:radial-exploded}, is called Radial Exploded layout. 
Selecting a specific node, the analysts can focus on its acquaintances that are displayed by using a radial layout. 
The characteristic of this exploded strategy is that if focuses only on a specific suspect and puts into evidence its links.  

\section{Case Study} \label{results}
\subsection{Aim of the Experimentation}
The aim of the current section is twofold: first of all, in order to highlight the potential of \emph{LogAnalysis} and the features provided to forensic analysts by the adoption of this tool, we discuss more into detail a number of examples, including the applicability of some centrality measures discussed above in the assessment of the importance of actors in phone call networks, the application of visualization techniques to highlight patterns of interactions among individuals in the networks, etc.

In addition, we underline that \emph{LogAnalysis} has been already adopted by one of the authors during several real-world criminal investigations. 
To this purpose, in this section we also report some data about these cases.
In particular, we provide details regarding several small and large case studies (including the details about the datasets of phone call networks adopted during the investigations) in which \emph{LogAnalysis} has been adopted to obtain additional insights regarding the networks structure.
Finally, this section is instrumental to introduce some additional features, voluntarily not discussed before, in order to understand their usage in the context of a real investigation. 

We tested our tool against different datasets (reported in Table \ref{tab:datasets}), whose size was comprised between about 4 thousands mobile communications and 8 millions.

\begin{table*}[!htb] \footnotesize \centering
	\begin{tabular}{| l | c | c | c | c | c | c | c | c |}
	\hline \hline
		Case no. &	1	&	2	&	3	&	4	&	5	&	6	&	7	&	8	\\
	\hline \hline
	No. nodes				&	148	&	170	&	381	&	461	&	320	&	543	&	912	&	702	\\
	No. edges				&	204	&	212	&	688	&	811	&	776	&	1,229	&	2,407	&	1,846	\\
	No. log entries &	4,910	&	8,447	&	125,679	&	250,886	&	280,466	&	589,512	&	7,567,119	&	8,023,945	\\
	\hline	\hline
	\end{tabular}
	\caption{Datasets adopted during real forensic investigations using \emph{LogAnalysis}. We highlight that the number of entries of the log files grows at a very different rate with respect to the number of nodes and edges in the network. This is a typical feature of the criminal phone call networks. }
	\label{tab:datasets}
\end{table*}

One important feature we discovered in the criminal phone call networks is the growth rate. 
We found that, even though the number of entries in the log files grows, the corresponding size of the network grows more slowly.
In fact, the analysis performed by forensic investigators is focused on the study of the network related to individuals that are already suspected of being involved in criminal activities, or being part of criminal organizations (i.e. the \emph{clans}) or terroristic groups. 
This reflects in a network whose structure grows slowly and comprises a relatively small number of nodes/edges (with different weights) with respect to the number of phone calls reported by the log files.

In our real phone call network criminal investigations, usually the analyst started with the study of the \emph{ego networks} of the individuals already suspect, those whose involvement in the criminal activities has been previously proved.
The main goal of the analyst was to disclose additional information about the underlying criminal organization.
For example, one important task was to put into evidence other individuals, whose activity was suspect, in order to hypothesize their complicity with actors whose involvement in the criminal organization was ensured.
This step was fundamental because, by identifying a small number of additional possible suspects, it has been possible to proceed with other ``traditional'' investigation methodologies, which would be not possible (in terms of time and cost constraints) otherwise.

\subsection{Further Details and Simple Use Case}
In the following we discuss an example use case that describes the usage of \emph{LogAnalysis} during criminal investigations.
As introduced above, analyzed data represent the phone call network of individuals suspected of belonging to criminal organizations.
The period of analysis usually coincides with the commission of certain serious crimes.
The adoption of our tool is instrumental to prove that those criminal facts have been planned and committed by the considered suspects.

Upon request by judiciary authorities, mobile phone service providers release all data logs about a certain set of suspected actors to the police force.
After the import of phone call data in \emph{LogAnalysis}, the process of analysis may start with the visualization of the phone call network by using the force-directed layout.
This is helpful to get a picture of the phone call network and the connections of suspected actors among each other and with other external individuals (see \ref{fig:loganalysis}).
Unless the number of individuals exceed thousands of actors, which requires a manual process of filtering, we remark that our tool is able to provide with a graphical meaningful visualization of the phone call network.
One advantage of the force-directed layout is the possibility of easily identifying clusters of actors within the network.

In order to improve the visualization, it is possible to apply some simple filters.
For example, once the forensic investigator identifies an actor of interest, just clicking on it, \emph{LogAnalysis} highlights those individuals with which the given actor is connected to, and those with which it shares the most of the contacts.
In that case, the number of in-going connections represent the \emph{popularity} of a certain actors and the number of out-going connections represents its \emph{gregariousness}.
It is easily possible to identify who are the individuals with respect to this actor is a gregarious, and who are those of which he/she exercise any influence.

Double-clicking on a given actor, the layout manager exploits the force-directed radial layout (see Figure \ref{fig:radial}). 
In such away it is possible, not only to have a picture of all the contacts of a given actor, but also to highlight the intensity with which those communications occur.
In addition, it is possible to put into evidence the \emph{affiliation} of each actor to a given cluster, identified by different colors (see Figure \ref{fig:radial-exploded}).

\begin{figure}
	\centering
	\includegraphics[width=\columnwidth]{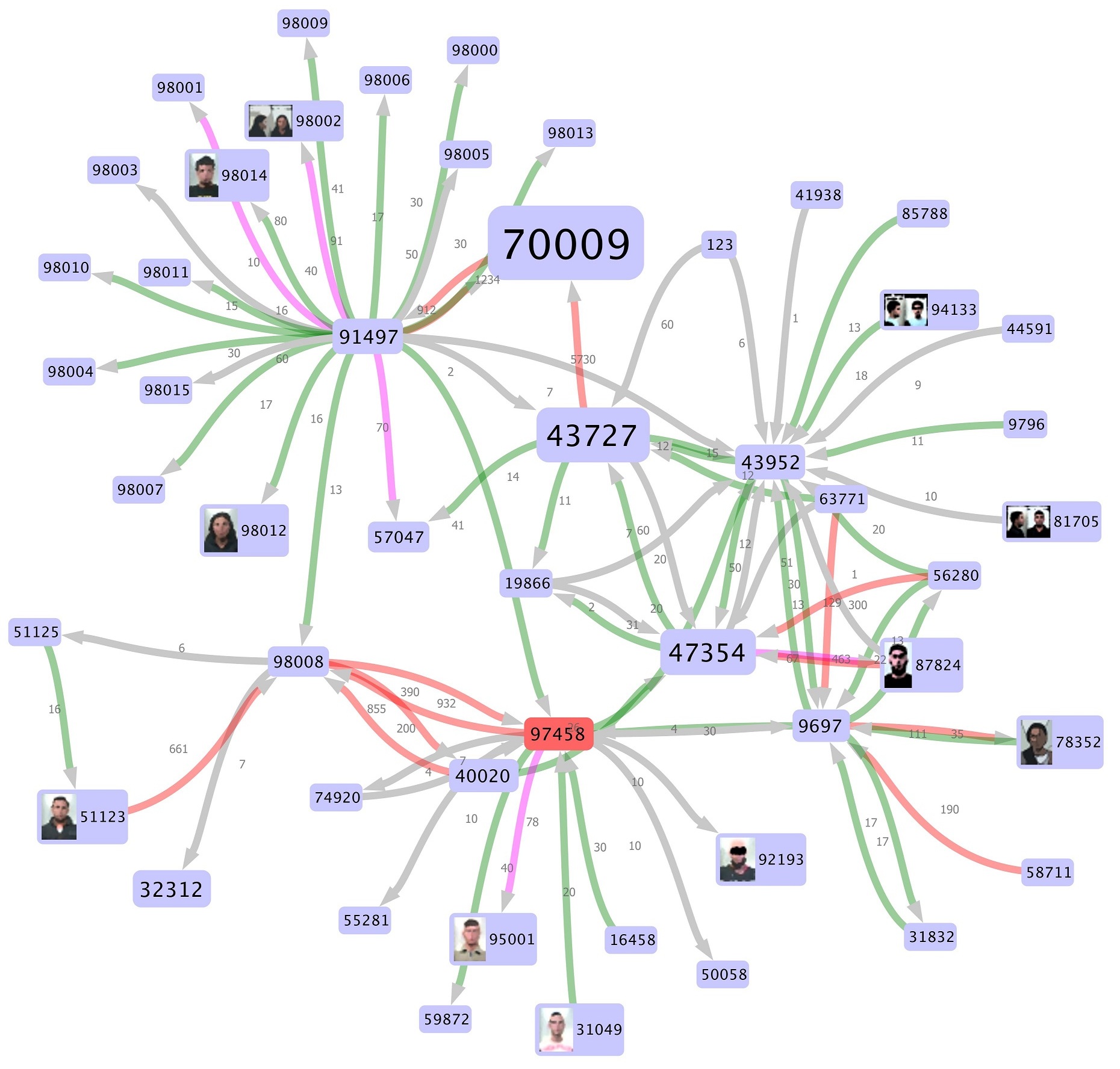}%
	\caption{Example of edge decorated. This feature classifies the edges with respect to a weight function and adopts different colors according to these weights. It is fundamental in order to give to the analyst an immediate picture that summarizes the intensity of the communications through the network.}%
	\label{fig:edge-decorated}%
\end{figure}

\emph{LogAnalysis} is particularly suited to assess the presence of clusters in the given phone call network and to visually put into evidence their structure (see Figure \ref{fig:visual-aggregation}).
This functionality is helpful to establish the role of a given set of actors inside a given group and to understand the structural and hierarchical organization of a possible criminal network.
To assess certain hypotheses on the hierarchical structure of given criminal network, the forensic analyst may exploit the tool depicted in Figure \ref{fig:edge-decorated}, that is helpful to have an immediate picture of the intensity of the communications among a set of actors, highlighting those connections whose relevance for the investigation is higher.

\subsection{Overall Metric Tool}
\begin{table*}%
	\footnotesize \centering
	\begin{tabular}{| l | c | c | c | c |}
	\hline	\hline
		Metric & Case 1 & Case 2 & Case 3 & Case 4 \\
	\hline	\hline
		No. log entries & 4,910	&	8,447	&	125,679	& 250,886 \\
		No. nodes & 148	&	170	&	381	&	461 \\
		No. edges & 240	&	212	&	688	&	811 \\
	\hline	\hline
		No. connected components & 1	&	1 & 1	& 1 \\
		Diameter & 6	& 7	& 7	& 6	\\
		Average geodesic & 3	& 3.418	& 3.898 & 3.514 \\
		Graph density & 0.002	& 0.010	&	0.005	&	0.004 \\
	\hline	\hline
		Minimum In-Degree	& 0	& 0	& 0	& 0	\\
		Maximum In-Degree	& 32	&	48	&	80	&	83 \\
		Average In-Degree	& 1.419	&	1.533	&	1.806	&	1.765 \\
		Median In-Degree	& 1	& 1 & 1 & 1 \\
	\hline	\hline
		Minimum Out-Degree	& 0 & 0 & 0 & 0 \\
		Maximum Out-Degree	& 33	&	38	&	78	&	81 \\
		Average Out-Degree	& 1.414	& 1.438	& 1.806	& 1.765 \\
		Median Out-Degree		& 1	& 1	& 1	& 1	\\
	\hline	\hline
		Minimum Betweenness Centrality	& 0 & 0 & 0 & 0 \\
		Maximum Betweenness Centrality (*)	& 12,975.267	&	14,345.581	&	63,244.345	&	85,132.261 \\
		Average Betweenness Centrality	& 358.932	&	443.231	&	1,105.139	&	1,154 \\
		Median Betweenness Centrality		& 0	& 0	& 0	& 0	\\
	\hline	\hline
		Minimum Closeness Centrality	& 0.001	& 0.001	& 0.001	& 0 \\
		Maximum Closeness Centrality	& 0.003	&	0.005	&	0.001	&	0.001	\\
		Average Closeness Centrality	& 0.002	&	0.003	&	0.001	&	0.001 \\
		Median Closeness Centrality		& 0.002	&	0.003	&	0.001	&	0.001	\\
	\hline	\hline
		Minimum Eigenvector Centrality	& 0 & 0	& 0	& 0	\\
		Maximum Eigenvector Centrality	& 0.077	&	0.004	&	0.033	&	0.040 \\
		Average Eigenvector Centrality	& 0.007	&	0.005	&	0.003	&	0.002	\\
		Median Eigenvector Centrality		& 0.494	&	0.376	&	0.001	&	0.001 \\
	\hline	\hline
		Minimum Clustering Coefficient	& 0	& 0	& 0	& 0	\\
		Maximum Clustering Coefficient	& 1 & 1	& 1	& 1	\\
		Average Clustering Coefficient	& 0.069	&	0.088	&	0.027	&	0.036	\\
		Median Clustering Coefficient		& 0	& 0	& 0	& 0	\\
	
	\hline	\hline
	\end{tabular}
	\caption{Overall metrics calculated on the datasets of our four case studies. (*) In \emph{LogAnalysis} the betweenness centrality is not normalized.}
	\label{tab:overall}
\end{table*}

In the following we are going to introduce additional features of \emph{LogAnalysis} that are instrumental in the context of real-world criminal investigations.

An important and useful feature provided by our tool discussed in this case study is related to the possibility of calculating global quantitative metrics on the nodes/edges of the network.
In particular, it is possible to evaluate some features usually adopted in SNA \cite{perer2006balancing} such as: 
i) overall network metrics (i.e., number of nodes and edges, density, diameter); 
ii) node rankings (i.e., degree, betweenness, closeness and eigenvector centrality) and, finally 
iii) edge rankings (by means of weights).

This first step has been helpful for the analyst to gain a first insight about the structure of the network, in particular putting into evidence individuals whose centrality values were suspect, with respect to the others.
Similarly, the same quantitative evaluation puts into evidence those connections (i.e., phone communications) that occur more frequently and those actors that are more active in the network.
In Table \ref{tab:overall} we report all the metrics calculated on the case studies 1-4.

\subsection{Data Visualization}
Figure \ref{fig:loganalysis}, \ref{fig:radial} and \ref{fig:radial-exploded} show some details about the \emph{LogAnalysis} user interface.
Once imported, data about the phone traffic network are visually represented by using the default view (i.e., by means of the aforementioned force-directed layout). 
Each node represents a cell phone, and edges indicate communications among them.
On the left, a control panel provides tools and filters in order to tune the visualization of the network.

Using the available dynamic filters, it is possible to hide or highlight those nodes (or connections) which satisfy specific criteria.
Moreover, analysts could interact with the graph, for example moving, hiding or emphasizing specific elements, in order to dynamically re-arrange the structure of the graph.

The visualization algorithm adopts a weighted representation of edges, drawing those edges with higher weights by means of thicker lines.
Standard nodes are represented by using light-blue as default color.
Color filters could be defined by users, accordingly to specific conditions.
For example, in this case study, ``light-green'' nodes reflect the ``arrested'' condition, ``light-red'' nodes accord to the ``sub tree control'' filter.

All these tools are provided in order to produce more readable network graphs.
It is additionally possible to adopt ``distance'' filters, excluding from the visualization all the nodes far from the selected one more than the user-defined threshold.
This is particularly helpful if the network that is under investigation is very large, constituted by more than thousands elements.

The optimal network visualization is a combination of both manual and automatic arrangements.
First, it is possible to automatically pan and zoom, so as the whole network fits the display area (this particular approach may or may not be appropriate, depending on the size of the network or on the specific task the analyst would like to perform).

However, the display automatically pans when a new node is expanded, centering on the newly expanded network. 
Additionally, the tool provides manual panning and zooming features in order to better satisfy user needs.
Moreover, it is possible to choose which kind of ``labels'' should be visualized, amongst the ``node-id'', the picture (if available), or both.
Even if the case study presented in this paper is inspired by a real investigation, for privacy reasons in the figures displayed in this work the pictures are fictitious. 
In the real investigations, these pictures represent the mugshots of suspects (for those who are available).

\subsubsection{Edge Decorated}
In this section we focus the attention on a specific feature regarding the visualization, called \emph{edge decorator}.
This technique we propose is optimal in the case both of phone traffic networks and, in our opinion, more widely in Social Network Analysis.

In detail, this strategy that has been introduced in \emph{LogAnalysis} produces graphs not only according to the force-directed layout, but also by adopting different colors not only for nodes but even for edges.
To this purpose, we recall that the node color is given by the clan each node belongs to.
Instead, the edge color is calculated by means of a weight function (in our case, the number of calls between a pair of nodes).
Edges are annotated with weights associated to both directions (in- and out-degree).
The interval in which the weights lie is normalized, depending on the characteristics of the network.
However, this strategy results in the adoption an \emph{edge color code}, that in our case study has been calculated as follows: 
(i) gray for $weight < 10$; (ii) green for $10 \leq weight \leq 60$; (iii) fuchsia for $61 \leq weight \leq 100$; (iv) red for $weight > 100$.

The main advantage of introducing color code for nodes and edges is the possibility of easily identifying the  strongest relationships, among hundreds, or even thousands, nodes and edges.
During the real investigation, this feature has been proved to be helpful in order to give to the analyst a clear picture of the intensity of the communications among different actors of the network, with the only effort to give a overall glance on the network itself.
Finally, the possibility of visually putting into evidence those communications paths that occur more frequently with respect to the average is instrumental because it allows to highlight in a visual way those information provided by the \emph{overall metric tool}.

\subsubsection{Shortest Path Finder}
\begin{figure*}[!ht]
	\centering
	\includegraphics[width=2\columnwidth]{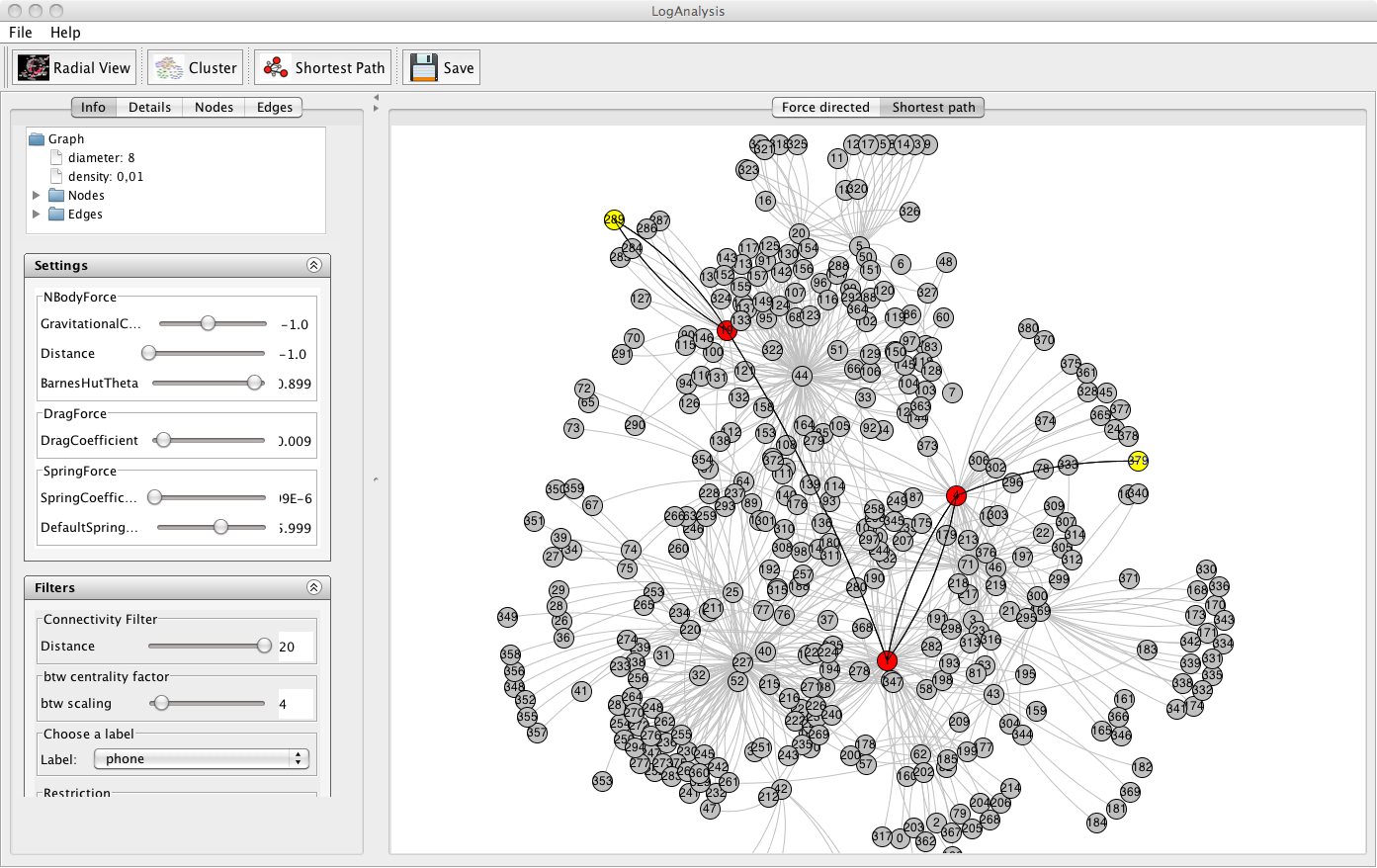}%
	\caption{The Shortest Path Viewer. It is fundamental when the analyst would like to understand the shortest ways of communications in the network. Usually, criminal organizations are structured in order to optimize the number of communications among members to efficiently disseminate information. This is possible by following short paths of communications that can be discovered by using this tool. }%
	\label{fig:shortest-paths}%
\end{figure*}

Another useful visualization tool provided by \emph{LogAnalysis} is the \emph{shortest path finder}.
The usage of the shortest path finder is crucial to highlight those paths that are optimal in order to spread information through the network.
In detail, the tool is useful to highlight nodes and edges involved in the shortest path between any given pair of nodes of the network.
This representation allows to highlight relationships among individuals belonging to distant groups in the graph.
In Figure \ref{fig:shortest-paths} the usage of the tool is presented.
In this specific case, the analyst was interested in understanding the most efficient way of communication that intervenes between nodes 289 and 379, two possible suspects.
Even though these nodes appears to be distant, it exists in the considered network a relatively short path, constituted only by 4 hops that connects these suspects.
Another essential information that it is possible to put into evidence by using this tool is that, usually, information can efficiently flow through those nodes that are more central in their respective clans, and that there exist usually a small number of \emph{referents} that vehiculate the most of the communications.

\subsection{Time Filtering} \label{sub:time-filtering}
\begin{figure*}[!ht]%
	\centering
	\includegraphics[width=2\columnwidth]{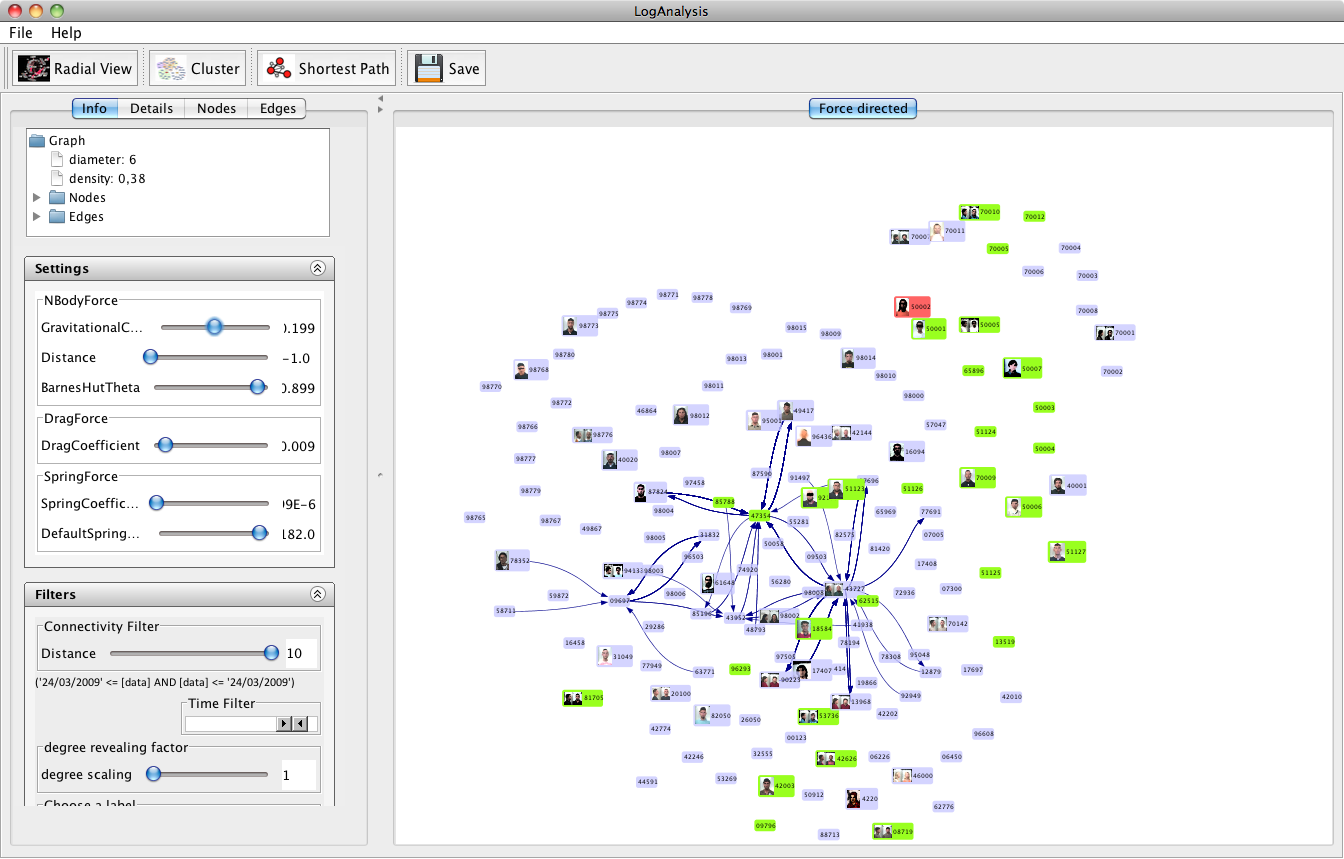}%
	\caption{The time filter feature. This tool is able to improve the capabilities of an analyst because it allows to specify a particular time window and to investigate how the structure of the network changes accordingly. Nodes are dynamically engaged or detached according to the time information about the phone call, dynamically altering the structure of the network. }%
	\label{fig:timefilter}%
\end{figure*}

A powerful filter included in \emph{LogAnalysis}, which deserves a specific explanation, is the \emph{time filter}.
Starting from the assumption that phone call networks are time-dependent, and the structure of the network could change accordingly, we introduced in our tool the possibility of ``filtering'' the structure of the network with respect to specific temporal constraints.
As shown in Figure \ref{fig:timefilter}, it is possible to select a time interval, by using a slider which comprises the whole temporal range covered by the log file.
The structure of the network is filtered accordingly, removing all the edges representing connections (i.e., phone calls) which did not take place in that specific time window, and insulating (or hiding) those nodes not involved in the network at that given time. 
Additionally, if the user modifies the time interval, nodes involved are automatically ``engaged'' or detached and, thanks to the force-directed algorithm, are attracted or rejected inside/outside the network.
The \emph{time filter} is a feature that has been proved to be incredibly powerful. 
Its adoption helps the analyst in identifying those communications that happened in a specific time window (say, for example, a day) and the structure of the graph during the given interval. 
Such a possibility heightens the capability of the detective to understand the structure of a criminal organization and its evolution over time.
In fact, because the connections may spread during a long time interval, it is fundamental for the investigator to understand at what time the given graph was already reflecting, for example, the structure of a clan or the presence of a particular \emph{referent} in the network.
Similarly, the possibility of dynamically visualize the effect of engaging or detaching nodes according to the modification of the time filter is crucial in order to highlight those nodes that are involved, during a specific time window, in the phone traffic network.

\subsection{Time Flow Analyzer} \label{sub:time-flow}
\begin{figure}
	\centering
	\includegraphics[width=\columnwidth]{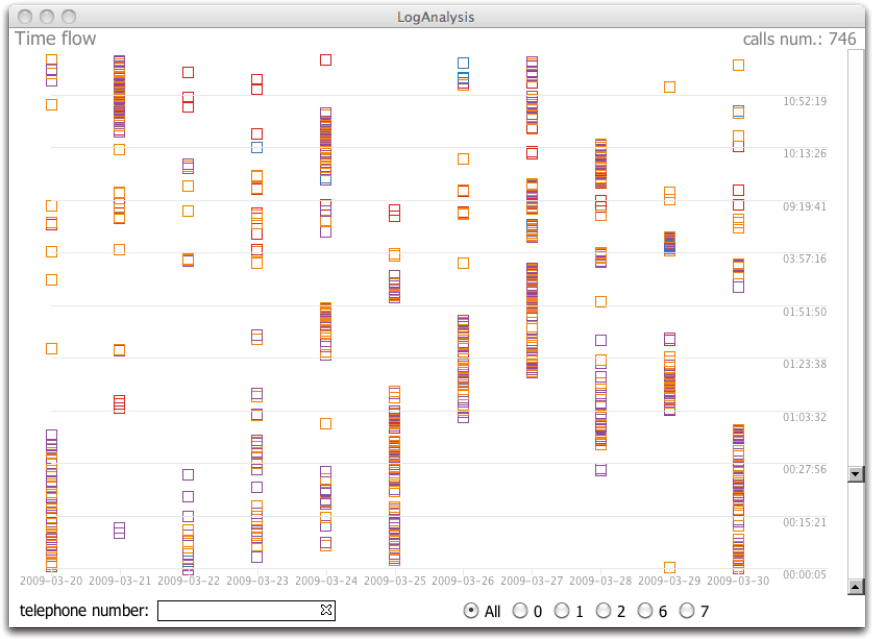}%
	\caption{The Time Flow Analyzer tool. This tool is helpful to consider the time-dependence of events (i.e., phone calls) in a specific time window and it is crucial in order to highlight phone call cascades during criminal events.}%
	\label{fig:time-analyzer}%
\end{figure}

The last visual tool which has been included in \emph{LogAnalysis} is related to the time filtering features previously presented, but it is also detached from the representation by means of a graph of the phone traffic networks. 
In fact, the \emph{Time Flow Analyzer} (see Figure \ref{fig:time-analyzer}) considers each single phone call as an \emph{event}, graphically represented in a time-line which covers a specific, user-defined, interval of time.
The advantage of a time-dependent visualization is crucial in the scenario of the forensic investigations.
In fact, it allows to organize information and event-flows in a visual manner in order to put into evidence the degree of correlation of specific events (in our case the phone connections).

In the \emph{Time Flow Analyzer} we included in \emph{LogAnalysis}, the visual representation of a bi-dimensional space presents the days on the x-axis and the hours on the y-axis.
Each event is presented by a colored square, whose color depends on the type of communication represented (i.e., sent/received calls and SMS and other type of communications, etc.).
It is possible to apply several filters, in order to select only specific events:

\begin{description}
	\item[All] All the phone events;
	\item[1-2] Sent/received calls;
	\item[6-7] Sent/received SMS;
	\item[0] All the other type of communications.
\end{description}

Moreover, it is possible to zoom in/out the time interval in order to obtain additional insights about connections of events.
Finally, the \emph{Time Flow Analyzer} allows the analysts to query the data in order to retrieve information about specific events or even about specific phone numbers, etc.
The adoption of this tool during real investigations is crucial to identify single events that set off to cascades of related events.
In detail, the time-dependent visualization allows the analyst to highlight those communications that triggered, in cascade, additional communications to other actors.
For example, it is possible to specify small time windows that may coincide to specific criminal events, in order to emphasize those phone connections that happened during the that interval and the involved actors, with an heightened probability of finding additional suspects or individuals involved in the criminal organization.

The aspect of temporal analysis in the context of phone call investigations has an extreme relevance.
The Time Flow Analyzer feature of \emph{LogAnalysis} allows to forensic analysts to highlight those fundamental communications that happened in critical periods of interest for a given investigation.
For example, from Figure \ref{fig:time-analyzer} it is possible to put into evidence that an important amount of phone calls happened before, during and after the commission of a serious crime, among those components of the criminal organizations highlighted by means of the network structure of the phone calls.
The temporal analysis, although not directly represented by means of networks, is closely interconnected to the structure and the evolution of the phone call network itself, and the Time Flow Analyzer tool is instrumental to highlight and understand this critical dependency.

\subsection{Stacked Histograms} \label{sub:stacked-histograms}
\begin{figure}
	\includegraphics[width=\columnwidth]{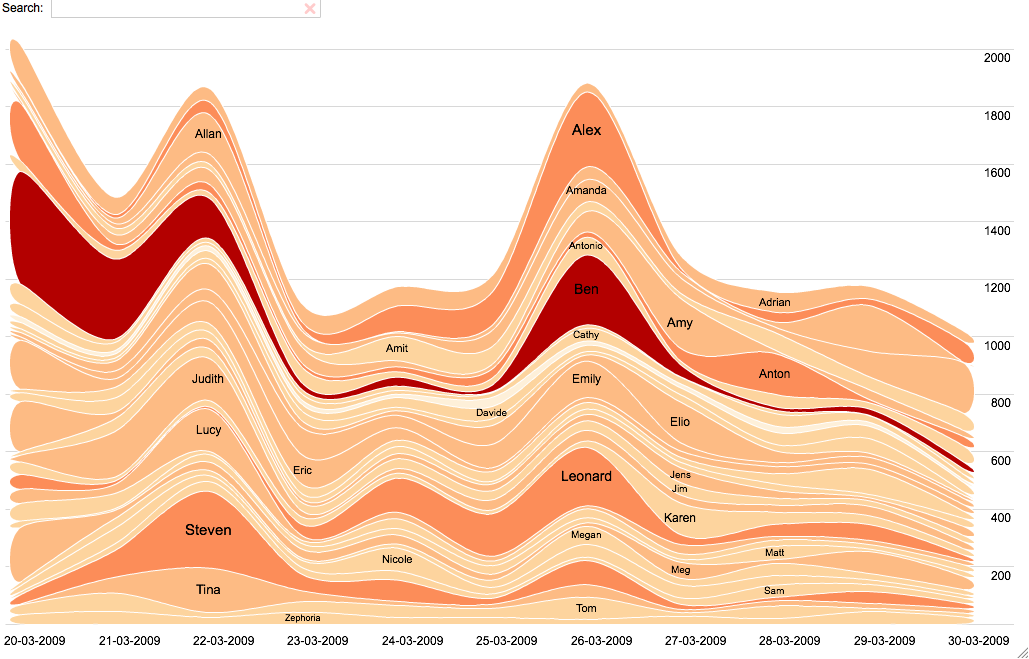}%
	\caption{The Stacked Histogram Tool is helpful to visually summarize the communications among actors elapsed in a given time interval.}%
	\label{fig:stream}%
\end{figure}

The last tool of \emph{LogAnalysis} described in this work is called Stacked Histograms. 
This tool empowers the temporal analysis features provided by \emph{LogAnalysis} and it is shown in Figure \ref{fig:stream}.
Its functioning is explained as follows.
Similarly to the Time Flow Analyzer tool, in the Stacked Histograms on the x-axis it is represented the time flow, but on the y-axis there is the amount of phone calls in the given interval.
In the Stacked Histograms, each actor has assigned a stack, whose color and intensity is proportional to the number of phone calls related to the given individual, during the specific period of interest taken in consideration from the forensic analyst.
In detail, the intensity of the color with which the stack histograms are represented is related to the absolute number of phone calls (in-coming and out-coming contacts) of each actor, while the thickness of the histogram may represent the in-degree or the out-degree of the given user at that day (highlighting those actors who are more popular and those who are more gregarious).
The Stacked Histograms tool is helpful to get a picture of the phone call activity of the set of considered actors elapsed during a specific time window.
Finally, it is particularly instrumental to understand in which proportion the phone activity of a given actor is with respect to the other individuals in its network who are in contact with him/her (i.e., its ego-network), in that specific time period.

\section{Conclusions}
The analysis of networks of phone traffic for investigative and forensic activities, aimed at discovering the relational dynamics among individuals belonging to criminal associations is a hard task.
Our goal was to develop a systematical model of analysis oriented to simplify exploration of networks whose elements are large collections of mobile phone traffic data.
Our approach is based on Social Network Analysis studies, which developed useful techniques to tackle the problem.
Nevertheless, few useful tools hitherto support this type of network analysis.
The tool we developed, \emph{LogAnalysis}, supports the exploration of networks representing mobile phone traffic networks.
It employs visual and statistical features in order to help in discovering cohesive groups, \emph{key figures} and individuals acting as link.
\emph{LogAnalysis} helps in systematically and flexibly obtaining measures typical of SNA in order to find outlier/anomalous values.
Users can interactively identify sub-groups and focus on interesting actors of the network.
In addition, the tool includes the possibility of exploring the temporal evolution of the network structure and the temporal information flow.

Future improvements to \emph{LogAnalysis} will concern the geo-spatial analysis of phone traffic networks and the implementation of novel measures of centrality \cite{ferrara2011novel,abdallah2011generalizing}, community detection algorithms and graph visualization techniques.

\begin{acknowledgements}
We would like to thank the Editor and the anonymous Reviewers whose comments helped us to greatly improve the quality of the work.
\end{acknowledgements}


\bibliographystyle{spmpsci}      
\bibliography{SNAM-2011}   

\end{document}